\documentclass[aps,preprint]{revtex4}
\usepackage{amsmath,amssymb,epsfig,bm}

\newcommand{\B}{\tilde{B}}
\renewcommand\d{\partial}
\newcommand{\Q}{\mathbf{Q}}
\newcommand{\bb}{\mathbf{b}}

\begin{document}

\preprint{INT-PUB 05-28}
\title{{\bf Anomalous Hydrodynamics}}
\author{G.~M.~Newman}
\affiliation{Institute for Nuclear Theory, University of Washington,
Seattle, Washington 98195-1550, USA}

\begin{abstract}
  Our goal is to examine the role of anomalies in the hydrodynamic
  regime of field theories. We employ methods based on gauge/gravity
  duality to examine R-charge anomalies in the hydrodynamic regime
  of strongly t'Hooft coupled, large $N$, $\mathcal{N}=4$ Super Yang-Mills.
  We use a single particle spectrum treatment based on the familiar
 ``level-crossing" picture of chiral anomalies to investigate thermalized
  massless QED. In each case we work in the presence of a homogeneous
  background magnetic field, and find the same result. Regardless of whether
  a particular current is anomalously non-conserved or not, as
  long as it participates in an anomalous 3-pt. correlator, its
  constitutive relation receives a new term: 
  \({\bf j}^{a} \propto - d^{abc}{\bf B}^{b}\rho^{c}\).
  This agrees with results found by Alekseev et.al. for QED.
  We include a general, symmetry based argument for the presence of
  such terms, and use linear response theory to determine their
  coefficients in a model with anomalous global charges.
  This last method, we apply to briefly examine baryon transport
  in chiral QCD in a strong magnetic field. 
\end{abstract}

\pagebreak

\maketitle

\section{Introduction}

Hydrodynamics \cite{hydro} provides a universal description of the long
range, long time-scale behavior of a wide variety of thermal systems.
The hydrodynamic quantities are those for which small perturbations
from equilibrium have relaxation times which diverge as their
wavelength diverges.  As a result, hydrodynamics has been formulated
as a classical effective field theory with various fields, typically
the conserved quantities, playing the role of fundamental degrees of
freedom.  Hydrodynamic behavior has been seen to arise from finite
temperature quantum field theory \cite{Jeon}.

At the same time however, one feature that sets aside relativistic
quantum field theories is the existence of quantum anomalies, which
cause some classically conserved quantities to be non-conserved.  An
unsolved problem is how to incorporate the effects of quantum
anomalies into the hydrodynamics description of a thermal field theory
which contain anomalies.  In a QCD plasma, the hydrodynamic theory
should reproduce the three-point correlation functions including the
anomalous part (an example of such correlation functions is that of
the axial vector, vector and baryon currents).  Another example is the
massless QED plasma (or the QED plasma at such high temperatures so
that the mass of the electron can be neglected).  In this case,
magnetohydrodynamics has to be enlarged to incorporate the axial
current $j^{5\mu}=\bar\psi\gamma^\mu\gamma^5\psi$.  Although this
current is not conserved, $\d_\mu j^{5\mu}\sim \bf E\cdot \bf B$, at
large distances the total axial charge should change slowly because the
conducting plasma cannot support a large long-distance electric field
$\bf{E}$.

In this paper, we consider a simpler problem where anomalies enter the
hydrodynamic equations already at the linearized level.  Namely, we
consider a theory with a set of global conserved charges $j^a$ for
which the triple correlators contain anomalous contributions.  We then turn on
an external magnetic field coupled to one of the global charges.  An
example of such a system is high-temperature QCD in a very large
magnetic field.  Now anomalies appear already at the level of
two-point correlators, and hence should be manifested in the
linearized hydrodynamic equations.

We shall argue that in the presence of a background magnetic field
coupled to some conserved charge, the constitutive equation for the
currents is modified in the presence of quantum anomalies.  In an
anomalous theory, any current $j^a$ participating in an anomaly now
receives a contribution
\begin{equation}
  {\bf j}^{a} \propto d^{abc}{\bf B}^{b}\rho^{c}.
\end{equation}
So, in addition to the diffusion and Ohmic current, there is a new,
dissipationless contribution proportional to the magnetic field 
and a density of charges. It is important to note that this modification 
of the constitutive relation occurs regardless of whether the current itself
is anomalously non-conserved. For example, in the massless QED mentioned
above, both the axial vector current and the electromagnetic vector
current receive this new term in their constitutive equations.
This result has been obtained previously, by Alekseev, Cheianov, and Frohlich, 
for the cases of 2D field theory, and massless QED~\cite{Alekseev}.
In this work, we will confirm their result, using different methods, 
as well as extending our analysis to a wider variety of systems.

We use two complimentary approaches to elucidate the impact of quantum
anomalies on the hydrodynamic regime.  The first approach uses the
gauge/gravity duality, in which the hydrodynamic behavior of currents
in 4D thermal gauge theories is obtained from the dynamics of the dual
Yang--Mills fields on black-brane backgrounds in higher-dimensions.
In this approach the 4D quantum anomaly has a very simple dual
description in the higher-dimensional theory --- it corresponds to the
5D Chern-Simons term in the gauge action.  This approach applies only
to strongly coupled gauge theories with gravity duals, in particular,
to ${\cal N}=4$ super-Yang-Mills theory in the large $N$, large
t'Hooft coupling regime.  In the second approach, which is appropriate
in the weak-coupling regime, we use a single particle spectrum treatment,
similar to the ``level-crossing'' pictures usually used for explaining
anomalies~\cite{Jackiw}, to examine the behavior of currents in a weakly coupled
abelian gauge theory with axial anomaly. In both cases we perform our
analysis in the background of a constant, homogeneous magnetic field
and obtain {\em the same} additional term in the constitutive relation
for an anomalous current.  We then argue that the form of the new term
in the constitutive equation that we found is exact, i.e., independent
of the strength of interactions.

The paper is organized as follows. In the first section, we use simple 
symmetry arguments to motivate the inclusion of terms of the form in~\cite{Alekseev} 
in the hydrodynamics of anomalous theories. In the second section, we explore
the R-charge anomaly in the hydrodynamic regime of $\mathcal{N}=4$ SYM
using a ``membrane paradigm'' treatment of the dual 5D theory which is
very similar to that used in~\cite{membrane} to examine non-anomalous
hydrodynamics in theories with gravitational duals. In the third section
we present a quasiparticle analysis of the axial anomaly in massless QED
at finite $T$ and finite (electron) chemical potential. Next we present
our argument for the universality of this result. This argument is based
on the equilibrium form of a charge distribution in a system with non-
vanishing gauge field. In section four, we consider the application of
our results to QCD at high temperature and strong magnetic field. 
Then, we conclude.

\section{Symmetry considerations}

The simplest model where the problem of anomalous hydrodynamics
appears is QCD with two massless quark flavors.  The conserved
currents in the theory are the isospin current $j^{a\mu}=\bar q
\gamma^\mu\frac{\tau^a}2 q$, the axial isospin current $j^{a\mu}=\bar
q \gamma^\mu\gamma^5\frac{\tau^a}2 q$, and the and the baryon current
$j_B^\mu = \bar q\gamma^\mu q$.  We turn on a background magnetic
field $\bf{B}$ coupled to the baryon current and discuss the
hydrodynamic behavior of the isospin and the axial isospin currents.

In the absence of the magnetic field $\bf{B}$ and at temperatures
higher than the chiral phase transition, the hydrodynamic equations
for the vector and axial charge densities $\rho^a$ and $\rho^{5a}$ are
the diffusion equations which can be written as the conservation laws
\begin{equation}
  \dot \rho^a + \bm{\nabla}\cdot{\bf j}^a = 0, \qquad
  \dot \rho^{5a} + \bm{\nabla}\cdot{\bf j}^{5a} = 0,
\end{equation}
coupled with the constitutive relations
\begin{equation}
  {\bf j}^a = - D\bm{\nabla} \rho^a, \qquad
  {\bf j}^{5a} = - D\bm{\nabla} \rho^{5a}
\end{equation}
We recall that the form of the constitutive equations are dictated by
the symmetries (rotational, C and P) and by the fact that we limit
ourselves, in linearized hydrodynamics, to terms linear in fields
with the lowest number of spatial derivatives.

However, in the presence of the external magnetic field $\bf{B}$, it
is possible to write additional linear terms
\begin{equation}
  {\bf j}^a = - D\bm{\nabla} \rho^a + c {\bf B}\rho^{5a},\qquad 
  {\bf j}^{5a} = - D\bm{\nabla} \rho^{5a} + c' {\bf B} \rho^a
\end{equation}
The equations are obviously rotationally invariant.  To see that they
respect C and P invariance one recalls that under C $j^\mu\to-j^\mu$,
$j^{5\mu}\to j^{5\mu}$, and ${\bf B}\to-{\bf B}$, and under P
$\rho\to\rho$, ${\bf j}\to-{\bf j}$, $\rho^5\to-\rho^5$,
${\bf j}^5\to{\bf j}^5$, and ${\bf B}\to{\bf B}$.  

The coefficients $c$ and $c'$ are not fixed by symmetries.  In the
next two Sections we compute these coefficients in some simple
theories.

\section{An approach from gauge/gravity duality}
\label{s-membrane}

\subsection{Introduction}

In this section we follow an analysis developed in
Ref.~\cite{membrane}, which employed the gauge/gravity duality and the
black-hole ``membrane paradigm'' to demonstrate hydrodynamic behavior
in a variety of finite-temperature theories with holographic
gravitational duals. We will work with the particular case of an
\(\mathcal{N} = 4\) super-Yang-Mills theory, believed dual to a stack
of black D3 branes in $AdS_{5} \times S^{5}$. (The hydrodynamic
behavior of this theory has been found directly.) For the sake of
completeness we will repeat many of the arguments used in
Ref.~\cite{membrane}.  This repetition also serves to demonstrate
explicitly where our assumptions do or do not differ from those of the
previous work.

The $AdS_{5}$ metric for our black brane configuration is,
\begin{equation}
  ds_{5}^{2} = \frac{(\pi TR)^{2}}{u}\biggl[-(1-u^{2})dt^{2}
  +\sum_{i=1}^{3}dx_{i}^{2}\biggr] + \frac{R^{2}}{4u^{2}(1-u^{2})}du^{2}. 
  \label{eq:metric}
\end{equation}
Note the use of the dimensionless radial coordinate, $u$, which is
related to the usual radial coordinate $r$ and horizon location
$r_{0}$ through \(u = r^{2}_{0}/r^{2}\).  The theory possesses $SU(4)$
conserved R-charges, whose correlation functions are computable using
AdS/CFT.  The the 5D bulk action of the gauge field $A^a_\mu$ dual to
the R-current, complete with Chern-Simons term is,
\begin{equation}
	S = 
	-\frac{1}{4g_{\rm GS}^{2}}\int d^{5}x\,\sqrt{-g} F_{\mu\nu}^{a}F^{\mu\nu a}
	-
	\frac{N^{2}-1}{96\pi^{2}}
	\int d^{5}x\,d^{abc}\varepsilon^{\mu\nu\lambda\rho\sigma}
	A_{\mu}^{a}\partial_{\nu}A_{\lambda}^{b}\partial_{\rho}A_{\sigma}^{c}. 
\end{equation}
In 5D, the Chern-Simons (CS) term is cubic in fields. Thus, two point
correlation functions can be calculated without this term, but three
and higher $n$-point functions receive anomalous contributions from the
CS term. The three-point functions can, in principle, be computed from
the closed-time-path AdS/CFT prescription. In this paper, we compute
two-point functions in the presence of a background magnetic field ---
which couples to the two-point functions through the CS term in the 5D
action,

In Ref.~\cite{membrane} the general idea, originated from the
black-hole membrane paradigm, begins with defining conserved currents
in terms of field tensors evaluated on a stretched horizon. The
prescription for how to define these currents is lifted directly from
the membrane paradigm~\cite{membrane_paradigm}. Then, one takes the long wavelength limit of
the field equations in order to derive the constitutive equation for
the currents. It was shown in Ref.~\cite{membrane} that the 5D Maxwell
equations imply Fick's law from boundary currents, with a diffusion
constant matching that found by direct AdS/CFT calculation in the case
of \(\mathcal{N} = 4\) SYM.

Working in the classical regime of the higher dimensional theory, we use the 
abelian field strength and ignore the $f^{abc}$ terms in the Yang-Mills action.
The modified Maxwell equations obtained from our action are,
\begin{equation}
   \frac{1}{g_{\rm SG}^{2}\sqrt{-g}}\,\partial_{\nu}[\sqrt{-g}F^{a \mu\nu}]+ 
   \frac{N^{2}\,d^{abc}}{128\pi^{2}\sqrt{-g}}
    \varepsilon^{\mu\lambda\nu\rho\sigma}F^{b}_{\lambda\nu}F^{c}_{\rho\sigma} 
   = 0 .
\end{equation}

In order to simplify our task, let us turn on a constant, homogeneous
background magnetic field, $\bf{B}$, in the 4D theory and discuss
charge diffusion on this background.  Turning on $\bf B$ in 4D means
imposing a boundary condition that the 5D field strength approach
$\bf B$ as $u \rightarrow 0$. One can see that these equations
support a constant magnetic field in the three space-like dimensions
perpendicular to the branes,
\begin{equation}
	-\frac{1}{2}\varepsilon^{ijk}F_{jk} = B^{i}.
\end{equation}
We will neglect the backreaction of this field on the background
metric. This is justified when $B \ll T^{2}$. Later on, we will work
in \(A_{5} = 0\) gauge and make use of a partial Fourier decomposition
of the $A_{\mu}$ into plane waves parallel to the horizon with
dependence on the fifth coordinate left explicit:
\[
  A_{\mu}(u,\omega,{\bf q}) = \int\,d^{4}x A_{\mu}(u,t,{\bf x})
  e^{i\omega\,t - i{\bf q}\cdot{\bf x}}.
\]
Throughout, we will use the notation
\[ 
  \B = \frac{B}{2(\pi T)^{2}} \:\:\:;\:\:\: \tilde{\omega} 
  = \frac{\omega}{2\pi T} \:\:\:;\:\:\: \tilde{q} = \frac{q}{2\pi T}. 
\]
In addition, we will make the assumptions \( \tilde{\omega} \sim
\tilde{q}^{2} \sim \B^{2} \) and \(\tilde{q} \ll 1\) which in the end
are seen to be consistent with the modified diffusion equation we
obtain.

The solution to the field equations describing a constant homogeneous
magnetic field, together with the metric (\ref{eq:metric}), defines
the 5D classical background about which we linearize in small
perturbations.  The linearized Maxwell equations become:
\begin{equation}	
     \frac{1}{g_{\rm SG}^{2}\sqrt{-g}}\,
     \partial_{\nu}[\sqrt{-g}g^{5\rho}g^{\nu\sigma}F^{a}_{\rho\sigma}] - 
     \frac{N^{2}\,d^{abc}}{16\pi^{2}\sqrt{-g}}B^{b}_{i}F^{c}_{ti} = 0, 
\label{eq:rad}
\end{equation}
\begin{equation}	
	\frac{1}{g_{\rm SG}^{2}\sqrt{-g}}\,\partial_{\nu}[\sqrt{-g}g^{t\rho}
        g^{\nu\sigma}F^{a}_{\rho\sigma}] + 
	\frac{N^{2}\,d^{abc}}{16\pi^{2}\sqrt{-g}}B^{b}_{i}F^{c}_{5i} = 0,
\end{equation}
\begin{equation}	
	\frac{1}{g_{\rm SG}^{2}\sqrt{-g}}\,\partial_{\nu}[\sqrt{-g}g^{i\rho}
        g^{\nu\sigma}F^{a}_{\rho\sigma}] - 
	\frac{N^{2}\,d^{abc}}{16\pi^{2}\sqrt{-g}}B^{b}_{i}F^{c}_{5t} = 0.
\end{equation}

In a theory with anomaly, the definition of current becomes ambiguous.
We shall require that all currents be gauge invariant under the
abelian subgroup of $SU(4)$ singled out by the magnetic field. Thus,
we employ the regularization where anomalies appear only in R-charge
currents for which no external gauge field is turned on. In this
regularization the divergence of an anomalous current is~\cite{Freedman:1998tz}:
\begin{equation}
\partial_{\mu}j_{a}^{\mu} = d^{abc}\frac{N^{2}-1}{128\pi^{2}}
  \varepsilon^{\alpha\beta\gamma\delta}
   F^{b}_{\alpha\beta}F^{c}_{\gamma\delta}. \label{eq:anom}
\end{equation}
In Ref.~\cite{membrane} the radial Maxwell equation, (\ref{eq:rad}) is
found to function as a conservation equation for the membrane paradigm
currents. In our case, we define currents for which the radial Maxwell
equation serves as the Adler-Bell-Jackiw anomaly equation,
(\ref{eq:anom}).  Ultimately, we will show that when we take
\begin{eqnarray}
	j_{a}^{t}& = &-\frac{\sqrt{-g_{5D}}}{g^{2}_{SG}}F^{5t}_{a}|_{u_{sh}}, \\
	j_{a}^{i}& = &-\frac{\sqrt{-g_{5D}}}{g^{2}_{SG}}(F_{a}^{5i} 
        - d^{abc}\B^{i}_{b}F_{c}^{5t})|_{u_{sh}}. \label{eq:constit}
\end{eqnarray}
to define the currents for our system, equation (\ref{eq:anom}) is
properly satisfied.  One can already see that when \(\bf{B} = 0\)
these currents are conserved by the Maxwell equation. The second term
on the right hand side of (\ref{eq:constit}) is a modification to
Fick's law, arising from the presence of the Chern-Simons term in the
action.

\subsection{Analysis}

We now embark on our study of linearized perturbations about our
solution. Our general strategy will be to demonstrate the same two
essential facts in~\cite{membrane}. First, as a direct result of incoming wave
boundary conditions at the horizon
\[ 
  F_{5i}(u_{sh})\propto F_{ti}(u_{sh}). 
\]
Second, in the hydrodynamic regime
\begin{equation}
  F_{ti}(u_{sh}) \approx -\partial_{i}A_{t}(u_{sh}). \label{eq:factII}
\end{equation}
Our analysis differs from \cite{membrane} - and not only by our inclusion of
new terms in the equations of motion and our accommodation of a
background $\bf{B}$ field. We also must employ different boundary
conditions on $A_{t}$ in order to allow for a non-vanishing $\bf{E}$
in the boundary theory. In \cite{membrane}, \(A_{t} \rightarrow 0\) at the
boundary leads, through (\ref{eq:factII}) to \(F_{ti} \approx
-\partial_{i}F_{5t}\), which supplies the diffusion term. In our case,
a non-zero $A_{t}(u=0)$ leads, in the same manner, to diffusion and
Ohmic terms in the constitutive relation, as well as an
$\bf{E}\cdot\bf{B}$ non-conservation term for an anomalous current.

We make use of three Maxwell equations (depending on whether the free
index is spatial, temporal, or radial) plus two Bianchi Identities:
\begin{eqnarray}
&&\partial_{t}F^{a}_{5t} - (1-u^{2})\partial_{j}F^{a}_{5j} 
  - d^{abc}\B^{b}_{j}F^{c}_{tj} = 0, \label{eq:Max-I} \\
&&\partial_{5}F^{a}_{5t} - \frac{1}{(2\pi T)^{2}u(1-u^{2})}
  \partial_{j}F^{a}_{tj} + d^{abc}\B^{b}_{j}F^{c}_{5j} = 0, 
  \label{eq:Max-II} \\
&&\partial_{5}[(1-u^{2})F^{a}_{5i}] - \frac{1}{(2\pi T)^{2}u(1-u^{2})}
  \partial_{t}F^{a}_{ti} - \frac{1}{(2\pi T)^{2}u}\partial_{j}F^{a}_{ij} 
  + d^{abc}\B^{b}_{i}F^{c}_{5t} = 0, \label{eq:Max-III} \\
&&\partial_{t}F_{5j} - \partial_{j}F_{5t} - \partial_{5}F_{tj} = 0, 
  \label{eq:Bnc-I} \\
&&\partial_{i}F_{tj} - \partial_{j}F_{ti} - \partial_{t}F_{ij} = 0. 
  \label{eq:Bnc-II}
\end{eqnarray}

As in \cite{membrane} a single wave equation for $F_{ti}$ can be obtained in
the near horizon limit.  Combining (\ref{eq:Bnc-I}) with
(\ref{eq:Max-III}) and (\ref{eq:Max-I}) respectively we find
\begin{equation}
	\frac{\partial_{t}^{2}F^{a}_{ti}}{(2\pi T)^{2}u(1-u^{2})} 
	- \partial_{5}[(1-u^{2})(\partial_{i}F_{5t}^{a} 
        + \partial_{5}F_{ti}^{a})]
	+ \frac{\partial_{j}\partial_{t}F_{ij}}{(2\pi T)^{2}u}
	- d^{abc}\B_{i}^{b}\partial_{t}F_{5t}^{c} = 0 ,             
        \label{eq:Wav-A}
\end{equation} \nopagebreak
\begin{equation}
	\partial_{t}^{2}F_{5t}^{a} - (1-u^{2})\partial_{j}
        (\partial_{j}F_{5t}^{a} + \partial_{5}F_{tj}^{a})
	 + d^{abc}\B_{j}^{b}\partial_{t}F_{tj}^{c} = 0 .
        \label{eq:Wav-B}
\end{equation}
Near horizon, these equations simplify significantly.  We proceed
under the assumption that all three terms of (\ref{eq:Max-I}) are of
the same degree of singularity as we approach the horizon, and check
the consistency of this assumption after the fact.  This allows us to
neglect $\partial_{j}^{2}F_{5t}$ in (\ref{eq:Wav-B}) and, passing to
momentum space, we find
\begin{equation}
	F_{5t}^{a} \sim -i(1-u^{2})\frac{q}{\omega^{2}}\partial_{5}F_{tj}^{a} 
	+ \frac{1}{\omega^{2}}d^{abc}\B_{j}^{b}\partial_{t}F_{tj}^{c} .  
        \label{eq:E5nh}
\end{equation}
It then follows that for \( 1-u \ll \omega^{2}/q^{2} \) the $F_{5t}$
terms in (\ref{eq:Wav-A}) can be omitted.  The Bianchi identity
(\ref{eq:Bnc-II}) indicates that the $F_{ij}$ term can be omitted as
well, and we obtain a wave equation for $F_{ti}$:
\begin{equation}
  \partial_{t}^{2}F_{ti} - (2\pi T)^{2}u(1-u^{2})\partial_{5}
  [(1-u^{2})\partial_{5}F_{ti}^{a}] = 0 .  \label{eq:Wav}
\end{equation}
Near horizon, this is solved by
\begin{equation}
	F_{ti}(u,t) = [\alpha_{i}(1-u)^{\frac{i\tilde{\omega}}{2}} 
	+ \beta_{i}(1-u)^{-\frac{i\tilde{\omega}}{2}}]e^{i\omega t} .
\end{equation}

If we specify incoming wave boundary conditions at the horizon, only
the first term can be allowed to contribute, and taking time and
radial derivatives of the last equation gives us
\begin{equation}
\partial_{t}F^{a}_{ti} - (4\pi T)(1-u)\partial_{5}F_{ti} = 0 .
\end{equation}
Bianchi identity (\ref{eq:Bnc-I}) then indicates, through (\ref{eq:E5nh}) that
\begin{equation}
\partial_{t}[F_{5i} - \frac{F_{ti}}{(4\pi T)(1-u)}] = 0 .
\end{equation}
Since finite energy solutions must decay with time, we have
\begin{equation}
F_{5i} = \frac{F_{ti}}{(4\pi T)(1-u)} .  \label{eq:B5nh} 
\end{equation}

Note that our results for $F_{ti}$ and $F_{5i}$ are consistent with
the assertion that all terms of (\ref{eq:Max-I}) are comparably
divergent as we approach the horizon. We have now established a
relationship between \( F_{5i} \subset j_{i} \) and $F_{ti}$. To
demonstrate a modified Fick's law, we must still relate this to a
gradient of $F_{5t}$. This is particularly straight forward in the
\(A_{5} = 0\) gauge, as \( F_{5t} = \partial_{5}A_{t} \). Below, we
have separeted $A_{i}$ into two pieces: \(A_{i} =
A^{\ast}_{i}({\bf x}) + A_{i}(u,t,{\bf x})\) where $A^{\ast}_{i}$
gives rise to the constant magnetic field, and $A_{i}(u,t,{\bf x})$ is
the arbitrarily weak perturbation about the classical background.  In
the hydrodynamic regime, and for a weak enough magnetic field, it is
possible to find solutions for $A_{t}(u)$ and $A_{i}(u)$,
perturbatively in $\tilde{\omega}$, $\tilde{q}$, and $\B$, such that
\begin{equation}
  \frac{A_{t}|_{sh} - A_{t}|_{0}}{\partial_{5}A_{t}|_{sh}} 
  \approx constant,   \label{eq:condA}
\end{equation}
while
\begin{equation}
  F_{ti}|_{sh} \approx -\partial_{i}A_{t}|_{sh}.   \label{eq:condB}
\end{equation}
Specifically, we take \( \tilde{q} \ll 1 \) and \( \tilde{\omega} \sim
\tilde{q}^{2} \sim \B^{2} \).

Finding the perturbative solutions to first order in the small
quantities is no simple matter for general $d^{abc}$, but it will not
be necessary for our purposes.  We will simply show that the leading
terms of both solutions satisfy the above conditions, determine the
value of the constant, and demonstrate that corrections to these
leading terms are small enough that the proceedure is valid.  As in~\cite{membrane} 
we take the streched horizon to be close enough to the actual
horizon to satisfy \( 1-u_{sh} \ll 1 \) without being exponentially
close:
\begin{equation}
  -\tilde{\omega}\ln(1-u_{sh}) \ll 1.
\end{equation}
Then, all the way down to the stretched horizon we can take
\begin{equation}
  (1-u)^{\frac{i\tilde{\omega}}{2}} \approx 1 .
\end{equation}

To isolate dependence on radial coordinate, we write the momentum
space Maxwell equations, using a prime to denote radial
differentiation:
\begin{eqnarray}
  &&\omega\,A^{a\prime}_{t} + (1-u^{2})q_{j}A^{a\prime}_{j} 
  + d^{abc}\B^{b}_{j}(q_{j}A^{c}_{t} + \omega A^{c}_{j}) = 0, 
  \label{eq:Mft-I} \\
  &&A^{a\prime\prime}_{t} - \frac{1}{(2\pi T)^{2}u(1-u^{2})}q
  (qA^{a}_{t} + \omega A^{a}_{j}) + d^{abc}\B^{b}_{j}A^{c\prime}_{j} = 0, 
  \label{eq:Mft-II} \\
  &&[(1-u^{2})A^{a\prime}_{i}]' + \frac{\omega\,}{(2\pi T)^{2}u(1-u^{2})}
  (qA^{a}_{t}+\omega A^{a}_{j}) \nonumber \\
  &&+ \frac{q_{j}}{(2\pi T)^{2}u}(q_{i}A^{a}_{j}-q_{j}A^{a}_{i}) 
  + d^{abc}\B^{b}_{i}A^{c\prime}_{t} = 0. \label{eq:Mft-III}
\end{eqnarray}

Then, solving equations (\ref{eq:Mft-II}) and (\ref{eq:Mft-III}) for
$\tilde{\omega} = \tilde{q} = \B = 0$ with the boundary conditions
\(A_{i}(0) = 0\) , \(A_{t}{0} = const.\) , \(A_{\mu}(u=1) = const.\)
we find
\begin{eqnarray}
  A^{a (0)}_{t}(u) = uC^{a (0)}_{t} + A^{a (0)}_{t}(0), \\
  (1-u^{2})A^{a (0) \prime}_{j} = C^{a (0)}_{j} .
\end{eqnarray}

Thus the constant relating $A_{t}$ to $\partial_{5}A_{t}$ is one. Now,
if we substitute (\ref{eq:B5nh}) into (\ref{eq:Mft-I}) (keeping in
mind we need only near horizon results to make statements concerning
the currents) we can relate $C^{a (0)}_{j}$ to $C^{a (0)}_{t}$.
\begin{equation}
  \omega A^{a\prime}_{t} + (1-u^{2})q_{j}A^{a\prime}_{j} 
  - i(2\pi T)d^{abc}\B^{b}_{j}[(1-u^{2})A^{c\prime}_{j}] = 0 .
\end{equation}
In fact, we have a matrix equation,
\begin{equation}
  A^{a\prime}_{t} = \frac{1}{\omega}[q_{j}\delta^{ac} 
  - i(2\pi T)d^{abc}\B^{b}_{j}][(1-u^{2})A^{c\prime}_{j}] ,
\end{equation}
that could be used to find $A^{(0)}_{j}$ from $A^{(0)}_{t}$ --- though
we need not do so here.  It is important to our argument that this
matrix equation be non-singular. While we are not aware of any
generally applicable reason it should not be, we can always restrict
our analysis to some subgroup of the R-charge $SU(4)$ for which this
will be true regardless of the relative value of $\tilde{q}$ and $\B$.
Thus, keeping to our assumption that \( \tilde{\omega} \sim
\tilde{q}^{2} \sim \B^{2} \), we have \( C^{a (0)}_{j} \sim
\frac{\omega}{q}C^{a (0)}_{t} \) and
\begin{equation}
  A^{a (0)}_{j} \sim A^{a (0)}_{t}\frac{\omega}{q}\ln(\frac{1+u}{1-u}) .
\end{equation}
Now we substitute \(A_{t} = A^{(0)}_{t} + A^{(1)}_{t}\), \(A_{j} =
A^{(0)}_{j} + A^{(1)}_{j} \) into equation (\ref{eq:Mft-II}), again
exploiting (\ref{eq:B5nh}), and find
\begin{equation}
  A^{a (1)\prime\prime}_{t} = 2\tilde{q}_{j}A^{a (0) \prime}_{j} 
  - d^{abc}\B^{b}_{j}A^{c (0) \prime}_{j} .
\end{equation}
Hence,
\begin{equation}
  A^{(1) \prime\prime}_{t} \sim \frac{\tilde{\omega}}{(1-u^{2})}A^{(0)}_{t},
\end{equation}
so,
\begin{equation}
  A^{(1)}_{t} \sim \tilde{\omega}A^{(0)}_{t} .
\end{equation}
Finally, invoking (\ref{eq:Mft-I}) again, we see 
\begin{equation}
  A^{(1)}_{j} \sim A^{(1)}_{t}\frac{\omega}{q}\ln(\frac{1+u}{1-u}) .
\end{equation}
Thus, we can safely conclude that \(F_{ti}|_{sh} \approx
-\partial_{i}A_{t}|_{sh}\).

\subsection{Summary}

In this way, we find ourselves in possesion of solutions satisfying
(\ref{eq:condA}) and (\ref{eq:condB}).  We now see that defining
currents as in (\ref{eq:constit}) allows (\ref{eq:Max-I}) to function
as the continuity equation. Application of (\ref{eq:B5nh}) now results
in a modified Fick's Law:
\begin{equation}
  {\bf j}^{a} = - D\bm{\nabla}\rho^{a} + \sigma\,{\bf E}^{a} 
  + d^{abc}{\bf \B}^{b}\rho^{c}.
\end{equation}
Here, $D=\frac{1}{2\pi\,T}$ is the diffusion constant, while
$\chi=\frac{N^{2}T^{2}}{8}$ is the susceptibility and
$\sigma=\frac{N^{2}T}{16\pi}$ the conductivity \cite{AdSCFT_hydro}.  As a
result, (\ref{eq:Max-I}) becomes a modified diffusion equation:
\begin{equation}
  \partial_{t}\rho^{a} - D\nabla^{2}\rho^{a} 
  + \sigma\bm{\nabla}\cdot{\bf E}^{a}
  + (\frac{N^{2}}{16\pi^{2}\chi})d^{abc}{\bf B}^{b}\cdot\bm{\nabla}\rho^{c} 
  = d^{abc}\frac{N^{2}}{16\pi^{2}}{\bf B}^{b}\cdot{\bf E}^{c}.
\end{equation}
As noted at the begining of this section, we cannot allow currents
coupled to a gauge field to be non-conserved lest the theory be
inconsistent. Hence, no one of the fifteen R-charges in the boundary
theory will have all of the above terms in its diffusion
equation. Charges coupling to a gauge field will have an Ohmic term,
but no ${\bf E}\cdot{\bf B}$ term:
\begin{equation}
\partial_{t}\rho^{a} - D\nabla^{2}\rho^{a} + \sigma\bm{\nabla}\cdot{\bf E}^{a}
+ (\frac{N^{2}}{16\pi^{2}\chi})d^{abc}{\bf B}^{b}\cdot\bm{\nabla}\rho^{c} = 0.
\end{equation}
 Anomalous charges will have no Ohmic term:
\begin{equation}
  \partial_{t}\rho^{a} - D\nabla^{2}\rho^{a}
  + (\frac{N^{2}}{16\pi^{2}\chi})d^{abc}{\bf B}^{b}\cdot\bm{\nabla}\rho^{c} 
  = d^{abc}\frac{N^{2}}{16\pi^{2}}{\bf B}^{b}\cdot{\bf E}^{c}.
\end{equation}

\section{Constitutive equations at weak coupling}

Above, we demonstrated an anomalous modification of constitutive
relations in a strongly coupled conformal field theory with a
gravitational dual.  We now turn to theories at weak coupling and
argue that the same modification should also appear there, and explain
its physical origin.

To keep the discussion simple, let's consider the theory a massless
fermion.  To have a hydrodynamic behavior at finite temperature the
fermion should interact with itself, but we shall assume the
interaction to be arbitrarily weak.  In accordance to the discussion
in the previous section, we turn on a background gauge field coupled
to the fermion current,
\[
  A^{\mu} = (0,0,Bx,0).
\]
Let us emphasize again that $A^\mu$ is only a background gauge field;
we do not include a dynamical U(1) gauge field into the theory.  The
fermionic Lagrangian for our theory is
\begin{equation}
  \mathcal{L}_{\psi} = \bar{\psi}\gamma_{\mu}D^{\mu}\psi,
\end{equation}
with $\psi$ a four-component Dirac spinor and \(D^{\mu} =
\partial^{\mu} - igA^{\mu}\).  The fermion Hamiltonian can be decomposed
into the left- and right-handed parts (Below, $\bm{\sigma}$
is the vector of Pauli matrices.)
\begin{equation}
H_{\psi} = i\int\,d^{3}x\,(\bar{\psi}_{\rm L}\bm{\sigma}\cdot{\bf D}\psi_{\rm L} 
                     - \bar{\psi}_{\rm R}\bm{\sigma}\cdot{\bf D}\psi_{\rm R}).
\end{equation}

Energy eigenstates of definite chirality can now be found which
satisfy \(-i\bm{\sigma}\cdot{\bf D}\psi_{\rm R} = E_{\rm R}\psi_{\rm R}\) and
\(i\bm{\sigma}\cdot{\bf D}\psi_{\rm L} = E_{\rm L}\psi_{\rm L}\). The solutions
can be separated via
\begin{equation}
  \psi_{\rm R,L}(x,y,z) = \left( \begin{array}{c} \varphi_{\rm R,L}(x) \\ 
  \phi_{\rm R,L}(x) \end{array} \right)e^{i(k_{y}y+k_{z}z)}.
\end{equation}
We have two pairs of coupled first ordered differential equations for
$\phi$ and $\varphi$ which can be written as a second order
differential equation for any one of them.  For example, 
\begin{equation}
  [\nabla^{2}_{\bar{x}} - e^{2}B^{2}\bar{x}^{2} + E^{2} - k_{z}^{2} + eB]
  \phi_{R} = 0.
\end{equation}
where \(\bar{x} = x - k_{y}/eB\),
Thus, we may write solutions for $\phi_{\rm R}$ as,
\begin{equation}
  \phi_{2}(\bar{x},n) = \left(\frac{eB}{(n!)^{2}2^{2n}\pi}\right)^{1/4}
  \mathcal{H}_{n}\left(\sqrt{eB}\bar{x}\right)e^{-\frac{eB}{2}\bar{x}^{2}}.
\end{equation}
Here, $\mathcal{H}_{n}(x)$ is the Hermite polynomial defined by
\begin{equation}
  \mathcal{H}_{n}(x) = (-1)^{n}e^{x^{2}}\frac{d^{n}}{dx^{n}}e^{x^{2}}.
\end{equation}
The energy spectrum is \( E_{\rm R}(n) = \pm\sqrt{k_{z}^{2} + 2neB} \),
where $n$ is the Landau level's label.  There is, however, a subtlety
at $n=0$. The equation from which $\varphi_{\rm R}$ must now be determined
is,
\begin{equation}
  [k_{z} - E_{\rm R}(n)]\varphi_{\rm R} = i(\nabla_{\bar{x}} + eB\bar{x})\phi_{\rm R}.
\end{equation}
When \(n = 0\) this equation can only be solved if \(E_{\rm R}(0) =
-k_{z}\). Thus, the energy eigenstates with $n=0$ are chiral: the
right-handed excitations form a single branch $E=-k_z$, while the
left-handed energy levels are $E=k_z$.

In the vacuum, all negative energy states are filled.  If one turns on
a chemical potential $\mu>0$ for the vector charge. then in addition
to the Dirac sea all energy levels with $0<E<\mu$ are populated.  For
the $n=0$ energy levels, this means left-handed fermions with
$0<k_z<\mu$ and right-handed fermions with $-\mu<k_z<0$ are populated.
But as all left-handed fermions have positive $k_z$ and right-handed
ones have negative $k_z$, the net result is that there is a nonzero
axial current that comes from the $n=0$ states.  One can also check
that the $n\neq0$ states do not contribute to the axial current, since
these energy levels are not chiral: the contribution from left- and
right-handed sectors cancel each other.  The effect survives at finite
temperature.

\begin{figure}[t]
\begin{center}
\def\epsfsize #1#2{0.6#1}
\epsffile{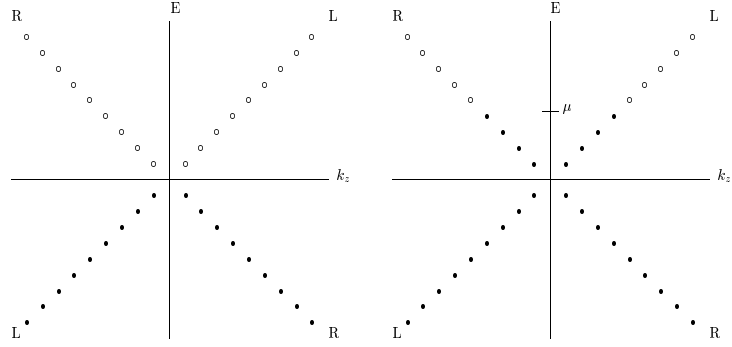}
\end{center}
\caption{A non-zero electric charge density gives rise to 
opposing fluxes of left and right handed particles,
culminating in \({\bf j}_{\rm R-L} \propto -\mu{\bf B}\).}
\label{fig:muEM}
\end{figure}

The effect can be quantified most easily by considering a finite
volume. If we place the theory in a square box of side length L, with
periodic boundary conditions, then the momentum runs discrete values
\(k_{i} = \frac2\pi n_{i}/L\).  Since the energy eigenvalues do not
depend on $k_{y}$, there is a degeneracy of various $k_{y}$ states at
each value of $E(n)$. Specifically, we requires \(\bar{x} = 0\) to be
inside the box forces \(0 \leq k_{y} < eBL\), which means that each
value of $n$ corresponds to $eBL^{2}/(2\pi)$ possible values of $k_y$.

In thermal equilibrium the total current is the sum over all energy
levels,
\[ 
  j^z_{\rm R,L}({\bf x}) = \frac1{L^3}
  \sum^{\infty}_{n=0}\sum_{k_z}
  \sum_{k_y}{E_{k}}f_{\rm R,L}(n,k_{z},k_{y},{\bf x}). 
\]
As mentioned above, states with \(n \ne 0\) do not contribute to the
axial current, since these positive energy states exist in equal
number with both signs of $k_{z}$, regardless of chirality.  The
function $f$ is the Fermi-Dirac distribution function,
\begin{equation}
  f_{R,L}(k_{z},k_{y};T,\mu) 
  =  \frac{1}{e^{\beta\,[|k_{z}| - \mu]} + 1}.
\end{equation}
After taking into account the degeneracy factor related to $k_y$,
summing over $k_{z}$, and regularizing by subtracting off the
contribution of the Dirac sea, one finds
\begin{equation}
  {\bf j}_5 = 
  {\bf j}_{\rm R}-{\bf j}_{\rm L} = -\frac{e}{2\pi^{2}}\mu{\bf B},
\end{equation}
which coincides with the result obtained for the gravity dual theory
in the previous section.

\begin{figure}[t]
\begin{center}
\def\epsfsize #1#2{0.6#1}
\epsffile{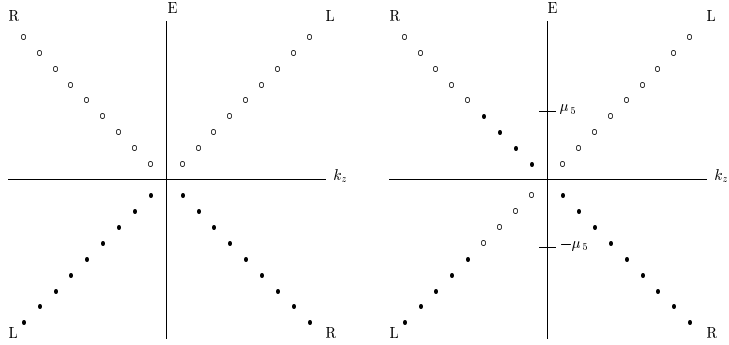}
\end{center}
\caption{A chiral density composed of equal numbers of 
right-handed particles and left-handed antiparticles gives
rise to a flux of positive charge in the direction of ${\bf B}$.}
\label{fig:mu5}
\end{figure}

The same analysis we just used can also be applied in determining the
electric current at equilibrium, in the presence of a non-zero
chemical for the axial charge
$\mu_5$ and homogeneous background ${\bf B}$. Just as in the case
of $\mathcal{N}=4$ SYM, we find that the constitutive relation of a
non-anomalous current is altered in the same manner as that of the
anomalous current. Specifically,
\begin{equation}
  {\bf j}_{\rm EM} = -\frac{e}{2\pi^{2}}\mu_{5}{\bf B}.
\end{equation}
(See Figure~\ref{fig:mu5}.)
This is the exact same result obtained in~\cite{Alekseev}.

\section{Coupling Independence}
\label{coup}

The presence of a new type of constitutive terms in the hydrodynamics of 
quantum field theories, on the background of a homogeneous magnetic field,
is to be expected on the basis of simple symmetry arguments.
These terms look like, \[ {\bf j}_{a} = \sum_{c}C_{abc}j^{0}_{b}{\bf B}_{c}\, . \]
In the case of a strongly t'Hooft coupled, large $N$ gauge theory dual to
an $AdS_{5}$ super-gravity, we see these terms arise naturally from
the R-charge anomaly in a membrane paradigm approach. In this strongly 
coupled case, the coefficients of the new terms are simply determined from 
the anomaly coefficient appearing in the divergence equation, and 
the susceptibility for the charge density in the new term. Specifically, 
\[ \d_{\mu}j^{\mu}_{a} = \eta\,d_{abc}{\bf E}^{b}\cdot{\bf B}^{c} \]
indicates that, \[ C_{ac} = -\frac{\eta}{\chi_{c}}d_{abc}\, . \]
(Since ${\bf B}$ is only present for one charge, $C$ need not depend on the index $c$.
Note also, that while the order of indices in unimportant in the tensor 
structure of the anomaly, it is none the less used in $C_{abc}$ to indicate
that we are considering the correction to the current, $a$, that is 
proportional to the charge density, $b$. The factor $\eta$ is dimensionless,
and geometric in origin.)
In the case of arbitrarily weakly coupled QED at finite temperature, we see
the exact same contribution to currents arise from a local thermal equilibrium
treatment, consistent with~\cite{Alekseev}. Here, the currents arise as a 
direct result of the effect of the chiral anomaly on the spectrum of the 
non-interacting theory. In this section we will formulate an argument as 
to why the coefficients of the new terms should generally be determined 
as they are in the two, disparate examples we have studied above.
We'll do this, by combining linear response theory with hydrodynamics,
and demanding that the hydrodynamics be consistent with the zero frequency,
zero momentum limit of a thermalized local quantum field theory.

We use linear response theory to express currents and densities in terms 
of retarded two point correlators, and perturbing source fields~\cite{Kapusta}. 
Then, we invoke symmetry arguments to build hydrodynamic equations for the 
currents. The hydrodynamic equations will then function as equations for the 
retarded correlators. Solving these equations and setting the energy to zero,
we find that enforcing the correct behavior of the retarded two point functions 
in the zero momentum limit constrains the coefficients of the constitutive terms 
to be exactly as they are in our two specific cases above. 
Doing that we assume the theory to contain no massless excitations that could 
cause the retarded two point functions to be divergent at exactly zero energy 
and small momentum. Specifically, this will mean that no Goldstone modes or 
massless, dynamical gauge fields can be present. The fact that the zero momentum 
limit enforces this constraint indicates that these "kinetic coefficients" 
for the new terms are actually determined by equilibrium physics, as observed in~\cite{Alekseev}.

We will work with a toy model containing two global charges, one 
vector and one axial-vector, participating in an anomaly with the same
background (vector) magnetic field. Because there are only two 
global currents participating in the anomaly in this system, we can
simplify our notation via $C_{\rm A}=C_{\rm AVV}$ and $C_{\rm V}=C_{\rm VAV}$.
The same analysis can be easily extended to a system with a more general 
set of currents. More will be said about this at the end of the section. 

LRT gives the response of a current to the presence of a source field, up to linear order
in the source, as
\[ \delta j^{\mu}_{a}(\omega,{\bf p})  = [\Pi^{\mu\nu}_{R}(\omega,{\bf p})]_{ab}A^{b}_{\nu}(\omega,{\bf p}).\]
Here, 
\[ [\Pi^{\mu\nu}_{R}(x-y)]_{ab} = \theta(x^{0}-y^{0})\langle[j^{\mu}_{a}(x),j^{\nu}_{b}(y)]\rangle, \]
is the retarded current-current correlator in the absence of the perturbation, and 
\[ \delta j^{\mu}_{a} = \langle j^{\mu}_{a}\rangle|_{A} - \langle j^{\mu}_{a}\rangle|_{0}, \] 
is the difference between the current's expectation value in the presence, and absence
of the source.
Here, we will employ a source for the temporal component of our vector
current only. For consistency with the rest of the paper, we will use 
the notation, \(\rho_{a} = \delta j^{0}_{a}\) in this section, thus:
\[ \rho_{\rm V}  = [\Pi^{00}_{\rm R}]_{\rm VV}A^{\rm V}_{0} 
\:\:\:\:\:;\:\:\:\:\: \rho_{\rm A} = [\Pi^{00}_{\rm R}]_{\rm AV}A^{\rm V}_{0}. \]

Substituting the LRT charge fluctuations into our modified diffusion equations gives,
\begin{eqnarray}
(\partial_{t} - D_{\rm V}\nabla^{2})\rho_{\rm V} &=& 
-\sigma_{\rm V}\bm{\nabla}\cdot{\bf E}_{\rm V} - C_{\rm V}{\bf B}_{\rm V}\cdot\bm{\nabla}\rho_{\rm A}, \\
(\partial_{t} - D_{\rm A}\nabla^{2})\rho_{\rm A} &=&
-C_{\rm A}{\bf B}_{\rm V}\cdot\bm{\nabla}\rho_{\rm V} + \eta d_{\rm AVV}{\bf B}_{\rm V}\cdot{\bf E}_{\rm V}.
\end{eqnarray}

Since \( {\bf E}_{\rm V} = -\bm{\nabla}A^{0} \), passing to momentum space
allows us to drop a factor of $\tilde{A}^{0}(\omega,{\bf p})$ from each
term in both equations, leaving us with equations for the current-current
correlators.
\begin{eqnarray}
(i\omega - D_{\rm V}p^{2})[\tilde{\Pi}^{00}_{\rm R}]_{\rm VV} &=& 
\sigma_{\rm V}p^{2} + iC_{\rm V}{\bf p}\cdot{\bf B}_{\rm V}[\tilde{\Pi}^{00}_{\rm R}]_{\rm AV}, \\
(i\omega - D_{\rm A}p^{2})[\tilde{\Pi}^{00}_{\rm R}]_{\rm AV} &=&
iC_{A}{\bf p}\cdot{\bf B}_{\rm V}[\tilde{\Pi}^{00}_{\rm R}]_{\rm VV} - i\eta d_{\rm AVV}{\bf p}\cdot{\bf B}_{\rm V}.
\end{eqnarray}
Solving for both correlators while keeping (for consistency)
only terms up to linear order in the magnetic field, we obtain
\begin{equation}
[ \tilde{\Pi}^{00}_{\rm R} ]_{\rm AV}  =  \frac{ iC_{\rm A}{\bf p}\cdot{\bf B}_{\rm V}[\sigma p^{2}-\frac{\eta}{C_{\rm A}}d_{\rm AVV}(i\omega-D_{A}p^{2})] }{ (i\omega-D_{\rm V}p^{2})(i\omega-D_{\rm A}p^{2}) },
\end{equation}
and,
\begin{equation}
[ \tilde{\Pi}^{00}_{\rm R} ]_{\rm VV}  =  \frac{ \sigma p^{2} }{ (i\omega-D_{\rm V}p^{2}) }.
\end{equation}

We are now equipped to ask what zero momentum behavior the hydrodynamic
equations instill in the retarded correlators, and to demand that this
be consistent with the known behavior from finite temperature field theory.
First, we will take $\omega\rightarrow 0$ and note that in this limit,
the retarded correlator becomes an analytic continuation of the euclidean time
(or Matsubara) two point correlator~\cite{Kapusta}. 
Now, taking $p\rightarrow 0$ we note that $\lim_{p\rightarrow 0}[ \tilde{\Pi}^{00}_{\rm R}]_{\rm AV}$ 
is well defined. To examine the momentum dependence of our hydrodynamic
$[ \tilde{\Pi}^{00}_{\rm R}]_{\rm AV}$ it will be convenient to separate the momentum 
into components perpendicular and parallel to the magnetic field. Writing 
\(B = |{\bf B}_{\rm V}|\) we find,
\begin{equation}
[ \tilde{\Pi}^{00}_{\rm R}(0,{\bf p})]_{\rm AV} = 
  (\frac{p_{\parallel}}{(p^{2}_{\parallel}+p^{2}_{\perp})})
  \frac{iC_{\rm A}B(\sigma+\frac{\eta}{C_{\rm A}}D_{\rm V})}{D_{\rm V}D_{\rm A}}.
\end{equation}
Observe, that if we take $p\rightarrow 0$ along a contour that keeps the ratio
$\frac{p^{2}_{\perp}}{p_{\parallel}}$ constant, the value of 
$\lim_{p\rightarrow 0}[ \tilde{\Pi}^{00}_{R}]_{\rm AV}$ will be entirely dependent
upon what the value of $\frac{p^{2}_{\perp}}{p_{\parallel}}$ is. Thus, the limit 
is not well defined, unless the numerator itself is zero. The only way this can 
happen in the context of our hydrodynamic equations, is for the following equality
to hold:
\[ C_{\rm A} = -\eta\frac{D_{\rm V}}{\sigma_{\rm V}}d_{\rm AVV}. \]
Taking the same limits for $\tilde{\Pi}^{00}_{\rm R}]_{\rm VV}$ returns the
susceptibility, as it should, and gives no information about $C_{\rm A}$ or $C_{\rm V}$.
The constant $C_{\rm V}$ can be found by repeating the same steps using only
a non-zero $A^{0}_{A}$, and examining $\tilde{\Pi}^{00}_{\rm R}]_{\rm VA}$.

As mentioned earlier, this analysis follows through
for more general sets of vector and axial-vector currents as well.
The currents are fluxes of global charges, of the form
\[
   j^{\mu}_{a} = \bar{\psi_{i}}\gamma^{\mu}V^{ij}_{a}\psi_{j} \:\:\:\:\:,\:\:\:\:\:
   j^{\mu}_{x} = \bar{\psi_{i}}\gamma^{\mu}\gamma_{5}A^{ij}_{x}\psi_{j},
\]
where $V_{a}$ and $A_{x}$ are the generators, in flavor space, of the symmetries 
giving rise to each charge. (We will use early Latin indices for vector current 
generators, and late Latin indices for axial current generators, for clarity in 
what follows.) In general, we can expect 
\( [V_{a},V_{b}] \ne 0 \), \( [A_{x},A_{y}] \ne 0 \), and \([V_{a},A_{x}] \ne 0 \).
The applied magnetic field couples only to one of the vector charges, which
we specify here by the index $b$.
In this case, symmetry will demand that new constitutive terms have the forms,
\[
  {\bf j}_{x} = \sum_{b}C_{xbc}\rho_{b}{\bf B}_{c} \:\:\:\:\:,\:\:\:\:\: 
  {\bf j}_{a} = \sum_{c}C_{azc}\rho_{z}{\bf B}_{c}.
\]
The equations of motion to be solved will now be matrix equations in the 
space of the global charges denoted by \{$x$,$a$\}.
Again, one finds that in order for the infrared limit of the two point
functions to behave properly,
\[ C_{xbc} = -\eta d_{xbc}\chi^{-1}_{b}, \]
must hold for each possible value of $x$, and $c$, while
\[ C_{azc} = -\eta d_{azc}\chi^{-1}_{z}, \]
holds for each $a$ and $z$, in the presence of ${\bf B}_{c}$. 
This line of argument applies regardless of the coupling strength 
of the theory under consideration.

\section{Quark Gluon Plasma in Magnetic Field}
\label{QGP}

It would be nice to get a look at the kind of effect our modified hydrodynamics
could have in a real physical system. One case in which transport behavior in 
a magnetized system with anomaly may be of interest is hot QCD in a strong
magnetic field~\cite{Cheng:1994yr},~\cite{Kabat:2002er}.

We will take a look at chiral QCD with three massless flavors, in the presence of
a homogeneous background $U(1)_{\rm EM}$ magnetic field.  This theory possesses
an axial anomaly coupling the electromagnetic ($Q$), baryon ($b$), and chiral EM ($5$)
currents. To be explicit, we write these currents in terms of the diagonal 
generators of $SU(3)_{f}$:
\begin{eqnarray}
   j^{\mu}_{Q}  & = & e\bar{\psi_{i}}\gamma^{\mu}\Q_{ij}\psi_{j} \\
   j^{\mu}_{5}  & = & \bar{\psi_{i}}\gamma^{\mu}\gamma_{5}\Q_{ij}\psi_{j} \\
   j^{\mu}_{b}  & = & \bar{\psi_{i}}\gamma^{\mu}\bb_{ij}\psi_{j} \label{cdef}
\end{eqnarray}
Where,
\[ \Q = \frac{1}{3}\left( \begin{array}{clcr}
                           2 &  0 &  0 \\
                           0 & -1 &  0 \\
                           0 &  0 & -1
                          \end{array} \right)
\:\:\:\:\:;\:\:\:\:\:
   \bb = \frac{1}{3}\left( \begin{array}{clcr}
                 1 &  0 &  0 \\
                 0 &  1 &  0 \\
                 0 &  0 &  1
               \end{array} \right) 
\:\:\:\:\:;\:\:\:\:\:
\psi = \left( \begin{array}{c}
               u \\ d \\ s
              \end{array} \right).
\]
There are two more currents diagonal in flavor which may be defined as
\[
j^{\mu}_{\lambda}  =  \bar{\psi_{i}}\gamma^{\mu}\lambda_{ij}\psi_{j} \:\:\:\:\:,\:\:\:\:\:
j^{\mu}_{\lambda 5}  =  \bar{\psi_{i}}\gamma^{\mu}\gamma^{5}\lambda_{ij}\psi_{j},
\]
with 
\[ \lambda = \frac1{2}\left( \begin{array}{clcr}
                     0 &  0 &  0 \\
                     0 &  1 &  0 \\
                     0 &  0 & -1
                    \end{array} \right). \]
These currents have hydrodynamic behavior, and are anomalous, but their hydrodynamic 
equations do not couple to the other currents, since \(Tr[\lambda\lambda\Q]=0\).
Thus, we ignore them.
The anomaly structure generates non-vanishing three point functions of the types,
$\langle j^{\nu}_{Q}j^{\lambda}_{Q}j^{\rho}_{5} \rangle$, and
$\langle j^{\nu}_{Q}j^{\lambda}_{b}j^{\rho}_{5} \rangle$.

In order that we be able to apply the methods of the last section, it is important 
that we take the system to be above the temperature at which the chiral phase 
transition takes place, so that no pion modes will be present 
to couple to the chiral current. Furthermore, we must treat the $U(1)_{EM}$
coupling as arbitrarily small, so the effects of dynamical gauge fields will be
negligible. Corrections due dynamic electromagnetic fields would be of order 
$\frac{\alpha_{\rm EM}}{\alpha_{s}}$.
Under these conditions, the electromagnetic current is diffusive, rather than ohmic.
Leptons are considered as absent. Thus, all three currents are carried only by quarks,
and will have identical diffusion constants.

If we apply the methods of the last section to this model, and
again work only to first order in the applied magnetic field, we find the following
set of coupled hydrodynamic equations. (To keep notation tidy, we will use 
\({\bf B}'\equiv\frac{e{\bf B}}{2\pi^{2}\chi_{\rm Q}}\) 
where ${\bf B}$ is the standard magnetic field.)
\begin{eqnarray}
(\d_{t}-D\nabla^{2})j^{0}_{Q} & = & -\frac{2}{9}{\bf B}'\cdot\nabla\,j^{0}_{5}, \\
(\d_{t}-D\nabla^{2})j^{0}_{5} & = & -\frac{2}{9}{\bf B}'\cdot\nabla\,j^{0}_{b}                                   
                                    -\frac{2}{9}{\bf B}'\cdot\nabla\,j^{0}_{\rm Q}, \\
(\d_{t}-D\nabla^{2})j^{0}_{b} & = & -\frac{2}{9}\frac{\chi_{\rm Q}}{\chi_b}{\bf B}'\cdot\nabla\,j^{0}_{5}.
\end{eqnarray}
We take the electric field to be zero here. Thus all three charges are conserved.
Passing to momentum space, we find a relatively simple eigenvalue equation for 
the dispersion relations, 
\[ \left( \begin{array}{clcr}
(i\omega - D_{b}q^{2})      &                  0               & -i\frac{2}{9}{\bf p}\cdot{\bf B}' \\
            0               &      (i\omega - D_{\rm Q}q^{2})      & -i\frac{2}{9}{\bf p}\cdot{\bf B}' \\
-i\frac{2}{9}\frac{\chi_{\rm Q}}{\chi_b}{\bf p}\cdot{\bf B}' & -i\frac{2}{9}{\bf p}\cdot{\bf B}' & (i\omega - D_{\rm Q}q^{2})
          \end{array} \right)\,
\left( \begin{array}{c}
         \rho_{b} \\ \rho_{\rm Q} \\ \rho_{5}
       \end{array} \right)
 = 0. \]
Clearly, for modes in which the momentum is perpendicular
to the magnetic field, the dispersion will be purely diffusive. 
This will also be the case, when \(\frac{B}{T^{2}} \ll \frac{q}{T}\),
regardless of orientation. Modes with momentum parallel to the magnetic 
field will exhibit more interesting behavior. Specifically we find the new 
(normalized) eigenmodes to be,
\begin{eqnarray}
\rho_{1} &\equiv& \sqrt{\frac{1}{1+\chi_{\rm Q}/\chi_b}}(-\frac{1}{\chi_{\rm Q}/\chi_b}\rho_{b}+\rho_{\rm Q}), \\
\rho_{2} &\equiv& \frac1{\sqrt{3+(\chi_{\rm Q}/\chi_b)}}( \rho_{b}+\rho_{\rm Q}+\sqrt{1+(\chi_{\rm Q}/\chi_b)}\rho_{5}), \\
\rho_{3} &\equiv& \frac1{\sqrt{3+(\chi_{\rm Q}/\chi_b)}}(-\rho_{b}-\rho_{\rm Q}+\sqrt{1+(\chi_{\rm Q}/\chi_b)}\rho_{5}), 
\end{eqnarray}
with dispersion relations,
\begin{eqnarray}
\omega_{1} &=& -iDp^2, \\
\omega_{2} &=& -iDp^2 - \frac{2}{9}\sqrt{1+(\chi_{\rm Q}/\chi_b)}{\bf p}\cdot{\bf B}', \\
\omega_{3} &=& -iDp^2 + \frac{2}{9}\sqrt{1+(\chi_{\rm Q}/\chi_b)}{\bf p}\cdot{\bf B}'.
\end{eqnarray}
Frequency eigenmodes are a mixture of currents defined in equation~(\ref{cdef}), 
and the eigenfrequencies now have a real component with the same form
as the dispersion of Alfven waves. (The asymptotic value of \(\chi_{\rm Q}/\chi_b = 2/9\)
may be substituted in, if exact numbers are desired.)
An interesting effect of this result, is that an over density of
$\rho_b + \rho_{\rm Q}$ will give rise to a dissipationless flow of this
charge along the direction of the magnetic field.

\section{Conclusions}

We have revisited the familiar ``level crossing'' picture of the
impact of a chiral anomaly on the non-interacting limit of a weakly
coupled gauge theory. There, we saw that the currents participating in
the anomaly are altered in a certain way, in the presence of a
homogeneous magnetic field: \({\bf j}^{a} = - \frac{e^{2}}{2\pi^{2}}{\bf B}^{b}\mu^{c}\).  
We have examined the hydrodynamics
of a strongly coupled (and strongly t'Hooft coupled) plasma using the
Gauge-Gravity duality as a tool. There, we saw the R-symmetry currents
of large $N$, $\mathcal{N}=4$ SYM receive a new term in their
constitutive relations, due to their participation in the anomaly:
\({\bf j}^{a} = - \frac{N^{2}}{16\pi^{2}\chi_{c}}d^{abc}{\bf B}^{b}\rho^{c}\).

There is a general symmetry argument for the inclusion of such terms in the
constitutive relations of any anomalous local quantum field theory. 
In the absence of massless modes coupling to the relevant currents,
we have a means of determining the coefficients of these terms by demanding
consistency between hydrodynamics and linear response theory in the 
zero momentum limit. Thus, we conclude that, in the presence of static magnetic
fields, the hydrodynamic 
constitutive relations of an anomalous QFT receive a contribution,
of the form $d^{abc}{\bf B}^{b}\rho^{c}$ with a coefficient of $\frac{1}{\chi_{c}}$
times a geometric factor that can be read off from an anomaly equation.
This holds regardless of coupling strength, and so long as their are
no massless dynamical modes coupling to the currents involved.

If this LRT treatment can be extended to systems with massless modes,
it would be possible to consider anomalous contributions to magnetohydrodynamics.
This is an interesting topic for future work, that may significantly broaden the
applicability of the considerations raised in this paper. 

This work is supported by DOE grant DE-FG02-00ER41132. The author wishes to thank 
D. T. Son, Pavel Kovtun, and Laurence Yaffe for useful comments and conversations.


\begin{thebibliography}{99}

\bibitem{hydro} 
Landau, Lifshitz, ``Fluid Mechanics",
London, Pergamon Press; Reading, Mass., Addison-Wesley Pub. Co., 1959.

\bibitem{Jeon}
S.~Jeon and L.~G.~Yaffe,
``From Quantum Field Theory to Hydrodynamics: Transport Coefficients and
Effective Kinetic Theory,''
Phys.\ Rev.\ D {\bf 53}, 5799 (1996)
[arXiv:hep-ph/9512263].

\bibitem{Alekseev}
A.~Y.~Alekseev, V.~V.~Cheianov and J.~Frohlich,
``Universality of transport properties in equilibrium, Goldstone theorem  and
chiral anomaly,''
arXiv:cond-mat/9803346.

\bibitem{Jackiw}
R.~Jackiw,
``Topological Investigations of Quantized Gauge Theories" (1983)

\bibitem{membrane}
P.~Kovtun, D.~T.~Son and A.~O.~Starinets,
``Holography and hydrodynamics: Diffusion on stretched horizons,''
JHEP {\bf 0310}, 064 (2003)
[arXiv:hep-th/0309213].

\bibitem{membrane_paradigm}
K.~S.~Thorne, R.~H.~Price and D.~A.~McDonald,
``Black Holes: The Membrane Paradigm," 
Yale University Press, New Haven 1986.

\bibitem{Freedman:1998tz}
D.~Z.~Freedman, S.~D.~Mathur, A.~Matusis and L.~Rastelli,
``Correlation functions in the CFT($d$)/AdS($d+1$) correspondence,''
Nucl.\ Phys.\ B {\bf 546}, 96 (1999)
[arXiv:hep-th/9804058].

\bibitem{AdSCFT_hydro}
G.~Policastro, D.~T.~Son and A.~O.~Starinets,
``From AdS/CFT correspondence to hydrodynamics,''
JHEP {\bf 0209}, 043 (2002)
[arXiv:hep-th/0205052].

\bibitem{Kapusta}
Joseph Kapusta,
``Finite Temperature Field Theory," 
Cambridge University Press, 1989.

\bibitem{Hosoya}
A.~Hosoya, M.~Sakagami, and M.~Takao
Annals Phys. {\bf 154}, 229 (1984).

\bibitem{LeBellac}
Michel Le Bellac
``Thermal Field Theory"
Cambridge University Press, 1996.

\bibitem{Cheng:1994yr}
B.~l.~Cheng and A.~V.~Olinto,
``Primordial magnetic fields generated in the quark - hadron transition,''
Phys.\ Rev.\ D {\bf 50}, 2421 (1994).

\bibitem{Kabat:2002er}
D.~Kabat, K.~M.~Lee and E.~Weinberg,
``QCD vacuum structure in strong magnetic fields,''
Phys.\ Rev.\ D {\bf 66}, 014004 (2002)
[arXiv:hep-ph/0204120].


\bibitem{SUSYhydro}
P.~Kovtun and L.~G.~Yaffe,
``Hydrodynamic fluctuations, long-time tails, and supersymmetry,''
Phys.\ Rev.\ D {\bf 68}, 025007 (2003)
[arXiv:hep-th/0303010].





\end{thebibliography}
\end{document}